\documentclass{aa}
\usepackage{graphbox}
\usepackage{graphicx}
\usepackage{txfonts}
\usepackage{epstopdf}
\usepackage{gensymb}
\usepackage{multirow}
\usepackage{color}
\usepackage{subfigure}
\usepackage{geometry}
\usepackage{lscape}
\usepackage[usenames,dvipsnames]{xcolor}
\usepackage{natbib}
\usepackage{amsmath}
\usepackage[utf8]{inputenc}
\graphicspath{{./Figures/}}

\begin{document}
\title{Doppler speeds of the hydrogen Lyman lines in solar flares from EVE}
\author{Stephen A. Brown\inst{1} \and Lyndsay Fletcher\inst{2} \and Nicolas Labrosse\inst{3}}
\institute{University of Glasgow, School of Physics \& Astronomy, Glasgow, G12 8QQ    \email{s.brown.6@research.gla.ac.uk} 
\and \email{Lyndsay.Fletcher@glasgow.ac.uk} 
\and \email{Nicolas.Labrosse@glasgow.ac.uk}}

\abstract
{}
{The hydrogen Lyman lines provide important diagnostic information about the dynamics of the chromosphere, but there have been few systematic studies of their variability during flares. We investigate Doppler shifts in these lines in several flares, and use these to calculate plasma speeds.}
{We use spectral data from the Multiple EUV Grating Spectrograph B (MEGS-B) detector of the Extreme-Ultraviolet Variability Experiment (EVE) instrument on the Solar Dynamics Observatory. MEGS-B obtains full-disk spectra of the Sun at a resolution of $0.1$nm in the range 37-105 nm, which we analyse using three independent methods. The first method performs Gaussian fits to the lines, and compares the quiet-Sun centroids with the flaring ones to obtain the Doppler shifts. The second method uses cross-correlation to detect wavelength shifts between the quiet-Sun and flaring line profiles. The final method calculates the ``center-of-mass" of the line profile, and compares the quiet-Sun and flaring centroids to obtain the shift.}
{In a study of 6 flares we find strong signatures of both upflow and downflow in the Lyman lines, with  speeds measured in Sun-as-a-Star data of around 10 km s$^{-1}$, and speeds in the flare excess signal of around 30 km s$^{-1}$.}
{All events showing upflows in Lyman lines are associated with some kind of eruption or coronal flow in imaging data, which may be responsible for the net blueshifts. Events showing downflows in the Lyman lines may be associated with loop contraction or faint downflows, but it is likely that  chromospheric condensation flows are also contributing.}


\maketitle

\section{Introduction}
Solar flares are a consequence of the coronal magnetic field's ability to store energy as it twists and becomes tangled. During a flare, the stressed magnetic field releases this energy by restructuring into a simpler configuration, enabled by magnetic reconnection. A large fraction of the liberated energy is then deposited in the chromosphere, resulting in the emission of radiation across the entire electromagnetic spectrum. The  standard model for flare energy transport from the tenuous corona to the dense chromosphere is by electrons which travel towards the flare footpoints, where they interact with and heat  the chromospheric plasma via Coulomb collisions \citep{Brown1971}.

Flares are dynamic events, capable of exceeding $10^{32}$ erg in energy output \citep{Fletcher2011}. Some of this energy goes into driving plasma motions and flows. Emission line spectroscopy can be used to measure plasma speeds, with line shifts thought to be due to Doppler shifted emission of moving material. For a flare on the solar disk, blueshifts and redshifts are interpreted, respectively, as bulk plasma upflows and downflows.

Flare upflows can be an indicator of chromospheric evaporation, an expansion of the chromospheric material due to the input of energy and subsequent heating of the plasma. In cases where the material cannot radiatively or conductively cool at a rate to balance the influx of heating, the evaporation is termed ``explosive" and is accompanied by downflows as a consequence of momentum-balance \citep{Canfield1987, Brosius2007, Milligan2009}. Generally, upflows are observed in high temperature lines (eg, \ion{Fe}{xiv-xxiv}) and downflows in cooler transition-region lines,  as confirmed in imaging spectroscopy observations  during explosive evaporation \citep{Milligan2009, Taroyan2014}. If the incoming flux of electrons is weaker than F$=10^{10}$ erg cm$^{-2}$ s$^{-1}$, evaporation will tend to be gentle and upflows can be observed even in low-temperature lines \citep{Zarro1988, Milligan2006, Battaglia2015}.

\begin{figure}
	\resizebox{\hsize}{!}{\includegraphics{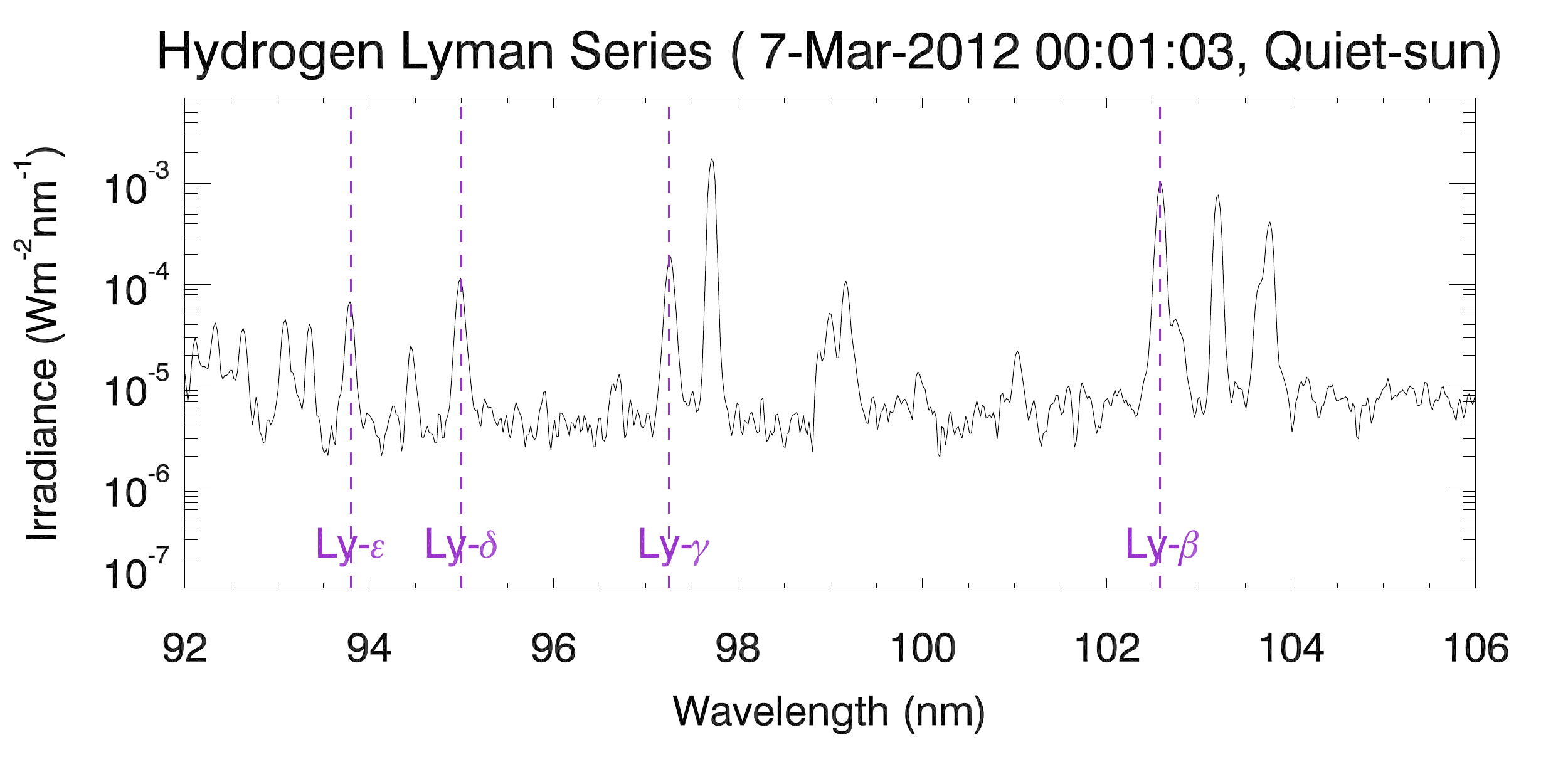}}
	\caption{Time-averaged spectrum during quiet-sun conditions of the hydrogen Lyman-series observed by MEGS-B just before the onset of the X5.4 SOL2012-03-07T00:07 flare.}
	\label{Fig1}
\end{figure}

We focus our study on the Lyman lines of hydrogen. These lines, formed by transitions to the $n=1$ state, lie longward of the Lyman continuum and edge ($\lambda_{edge}=91.2$ nm) and populate the EUV region up to Ly-$\alpha$ at 121.5 nm. They are well-observed by the Extreme Ultraviolet Variability Experiment (EVE) instrument on the Solar Dynamics Observatory (SDO) \citep{Woods2012}. EVE consists of several detectors; MEGS-A, which took data up to May 2014 , covers the $5-37$ nm range, while MEGS-B covers $35-105$ nm, both doing so at a spectral resolution of $0.1$ nm. MEGS-B operates on a reduced duty cycle, obtaining spectra continuously for 3 hours per day in addition to 5 minutes per hour. An example of a quiet Sun MEGS-B spectrum with the positions of Ly-$\epsilon$ through Ly-$\beta$  is shown in Figure \ref{Fig1}.

Only a few observations of the Ly-series during flares have been made. In early spectroscopic observations,  \citet{Lemaire1983} observed enhancements in the Ly-$\alpha$ and Ly-$\beta$ line profiles during a flare, along with a redshift in Ly-$\alpha$ corresponding to a speed of around 12 km s$^{-1}$. Similarly, \citet{Woods2004} observed enhancements in both the wings and the core of the Ly-$\alpha$ line. \citet{Kretzschmar2013} observed a small enhancement ($0.6\%$) in Ly-$\alpha$ during an M2 flare, while \citet{Milliganetal2014} found the line to be the dominant measured emission line in an X2.2 flare. Imaging  observations by the Transition Region and Coronal Explorer (TRACE) mission of Ly-$\alpha$-channel emission in flare footpoints were reported by \citet{Rubiodacosta2009}, who found the ratio of footpoint to quiet-Sun emission in Ly-$\alpha$ to be around 80. These observations also revealed a filament eruption in Ly-$\alpha$. This eruption, traveling at a projected speed estimated at $300$ km s$^{-1}$ had  less than 10\%  of the surface brightness of the chromospheric footpoints, but with a much larger emitting area and would likely have contributed to the flare Ly-$\alpha$ intensity

\section{Methods}~\label{section:methods}

\begin{figure}
	\resizebox{0.68\hsize}{!}{\includegraphics{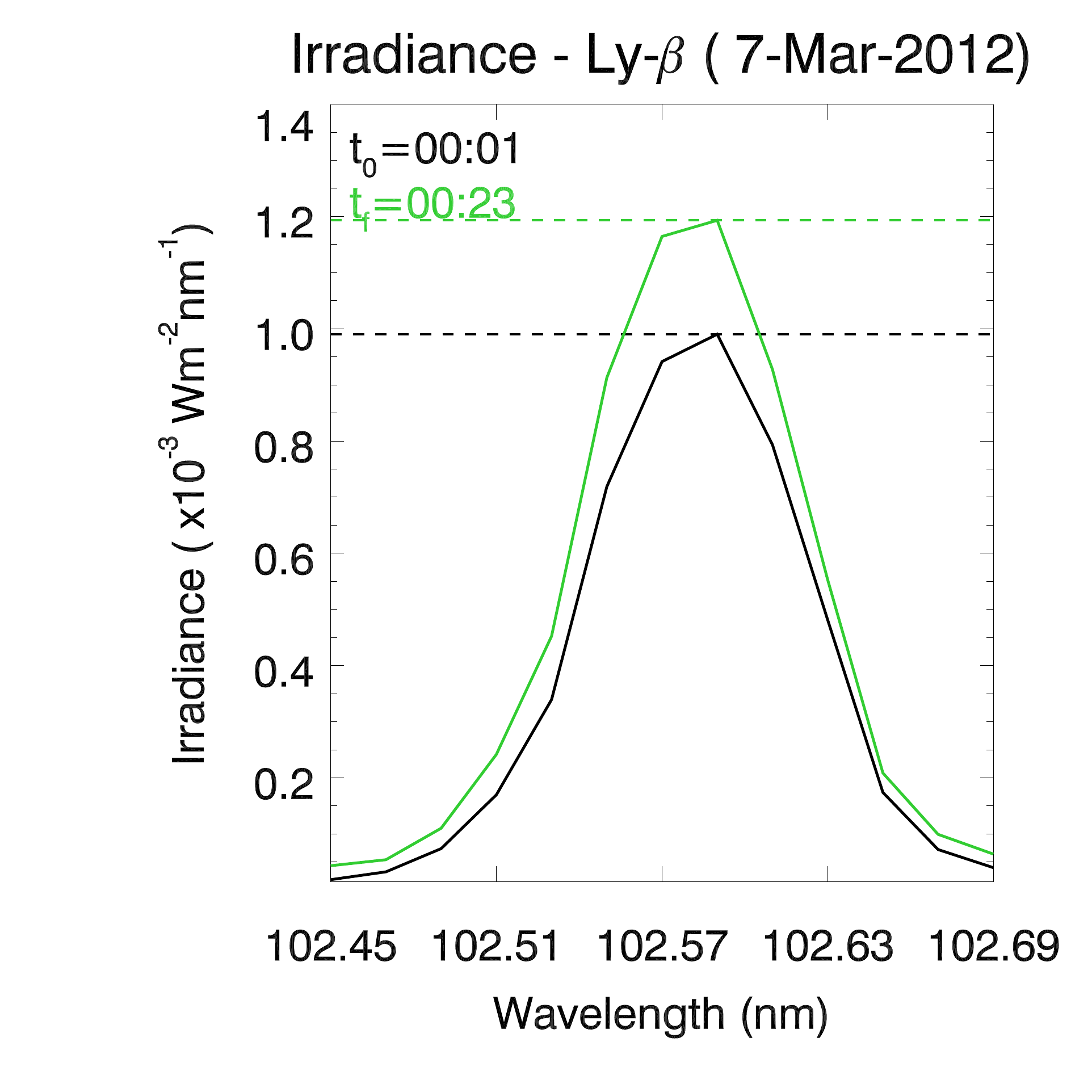}}
	\caption{The Ly-$\beta$ line before the onset of the 07th March 2012 flare (black) and during the flare (green). An irradiance enhancement can be seen in the core of the line, indicated by the position of the dashed line.}
	\label{Fig2}
\end{figure}

The MEGS-B data consist of hour-long data files containing spectra at a cadence of ten seconds and a spectral resolution of 0.1 nm. The wavelength-sampling is 0.02 nm \citep{Woods2012}. We use the EVE Level 2 Version 5 data in this study.  To identify our sample of flares, we first found all strong M and X flares from 2011 to early 2015  for which the impulsive phase and peak had been caught in the MEGS-B daily observing window. Ly-$\beta$  is a strong line in the quiet Sun, so we inspected the Ly-$\beta$ lightcurves of this initial sample of 17 flares to find flares with a good enhancement above the pre-flare background, in order to ensure the presence of a ``flare-excess" spectral signal in the Lyman lines.  The enhancement in the Ly-$\beta$ line from the peak of the strongest flare in our sample is shown in Figure \ref{Fig2}, showing that even in this event only a 20\% increase is seen. These selection criteria on the Ly-$\beta$ light curves resulted in 6 flares for further study. With the exception of SOL2011-03-07T19:46 (M3.7), they are all strong M and X class events (stronger than M9.9). We fitted the Ly-$\beta$ line in all initial 17 events, but found that systematic, detectable line shifts were also only present in the 6 strong M and X events. This seems reasonable, since a good signal-to-noise ratio is required to make the Doppler measurements, and we would expect the most dramatic motions and strongest Doppler shifts to be present in intense flares, either through explosive evaporation/condensation or mass motions. 

The EVE instruments perform Sun-as-a-star observations, meaning that measurements of the irradiance of a spectral line will contain contributions from across the entire disk,  including flaring chromosphere, non-flaring chromosphere and also flare-related filament eruptions such as those reported by \citet{Rubiodacosta2009} which may be emitting in low-temperature lines, or scattering it from the chromosphere.
 In order to isolate the emission from the flaring active region, we need to remove the pre-flare contribution. We do this by time-averaging as much of the spectrum as is available before the flare initiates, and subtracting it from the spectra during each of the flaring times. This gives us two possible data types to work with; the ``Sun-as-a-star" spectra with no background subtraction, and the ``flare-excess" spectra which have had the pre-flare background subtracted. 

 We include disk-positions for the flares in Table \ref{Table1}. However, we do not factor the location of a flare into our analysis. Even assuming flows aligned purely along the chromospheric magnetic field, this field needs not be locally vertical to the surface, so near-limb flares may still produce an observable (albeit, weak) Doppler shift. If we restricted our assumptions to purely-vertical field alignments then it should still be possible to observe Doppler speeds along the line-of-sight in all of our flares, as the maximum flare longitude in our sample is $64\degree$, giving a line-of-sight component of $0.44$ that of the overall velocity.

In order to quantify the intrinsic noise (due to solar variations, photon counting statistics and instrumental noise) in the EVE irradiance measurements, we calculate the standard error on the mean of the irradiance of the central Ly-$\epsilon$ wavelength bin, the mean being obtained from averaging the irradiance values in the bin throughout the time range defined as during ``quiet-Sun" conditions. This particular wavelength bin was chosen because the higher-order Lyman lines are situated in an increasingly crowded part of the spectrum, and represents a conservative estimate on the variability of the irradiance. We use three different methods to calculate the Doppler shifts, as described below.

\subsection{Method 1 - Gaussian fitting}
\subsubsection{Single Gaussian fitting}
The flare's lightcurve in Ly-$\beta$ is examined to ascertain the beginning and end times of both the quiet-Sun and flaring periods. These are used to define the time-range to average the pre-flare spectra over, and the timerange for which to calculate the flaring Doppler shifts. The amount of pre-flare data available depends on how close the flare's start-time is to the start-time of the data. In some cases, there are only a few minutes of pre-flare spectra available.

\begin{figure}
	\resizebox{0.85\hsize}{!}{\includegraphics{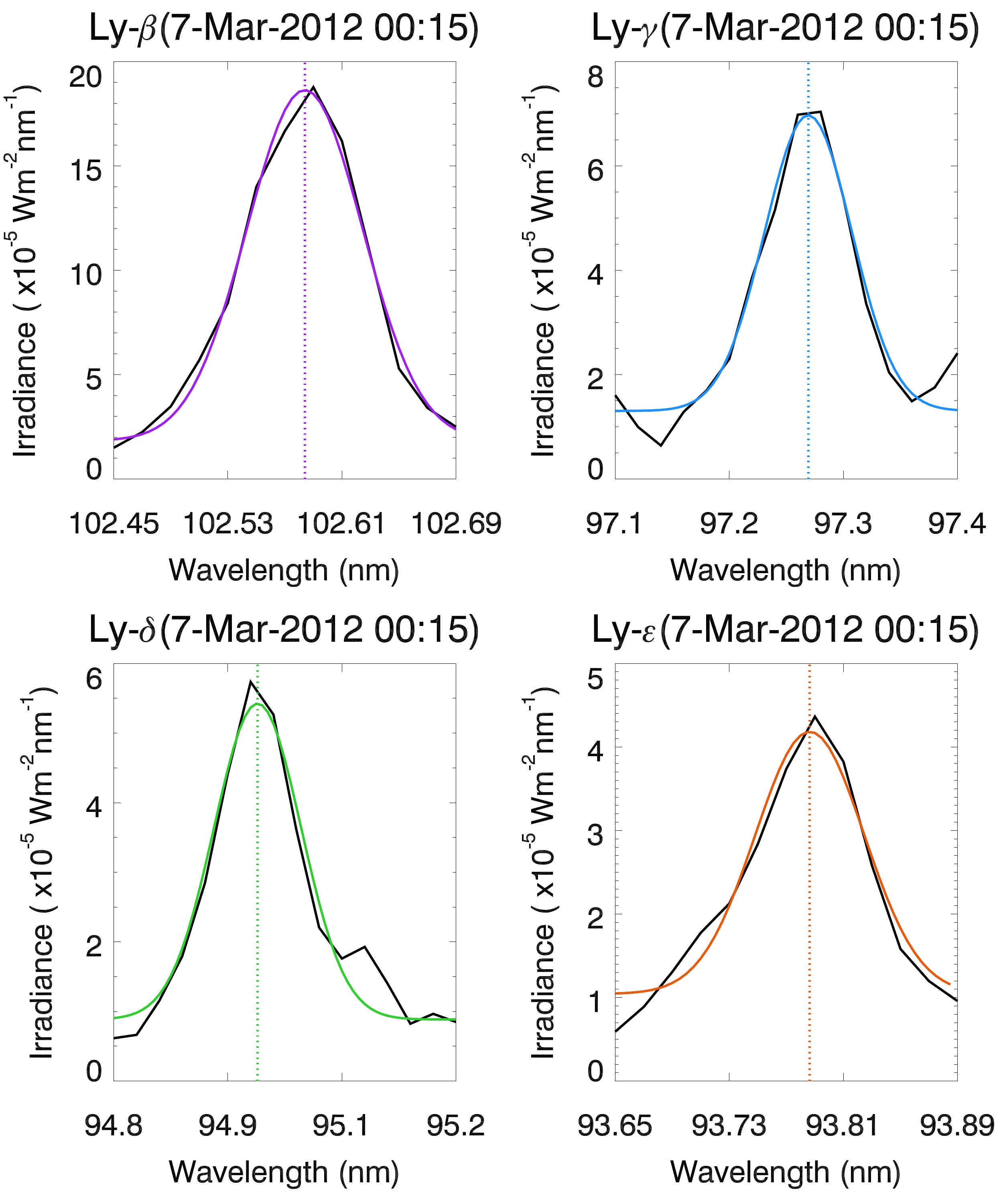}}
	\caption{Gaussian fits to the flare-excess Lyman lines during one of the $10$ s integrations on the 07th March 2012 flare. Some noticeable asymmetries can be seen in the line profiles.}
	\label{Fig3}
\end{figure}

For ``flare-excess" measurements, the pre-flare irradiance values are subtracted from flaring spectra. An IDL least-squares fitting procedure (``mpfitexpr.pro", \citet{mpfitfun}) is used to perform a Gaussian fit with a constant background over each of the line profiles, which then returns their mean wavelengths. The pre-flare spectrum's line profiles are also fitted in this way, giving us the quiet-Sun line mean wavelengths. Examples of the Gaussian fits can be seen in Figure \ref{Fig3}. For each flaring time, a Doppler shift can be calculated by subtracting the quiet-Sun mean wavelengths from the flare-excess ones. These shifts are then converted into speeds using the standard Doppler formula. If the ``Sun-as-a-star" approach is taken, the process is the same with the exception that no pre-flare background subtraction takes place. Errors are calculated by the ``mpfitexpr" procedure using the irradiance errors described earlier. This gives us formal error estimates for each of the parameters, in particular the line-centroid position that concerns us.

\subsubsection{Double Gaussian fitting}
\begin{figure}
	\resizebox{\hsize}{!}{\includegraphics{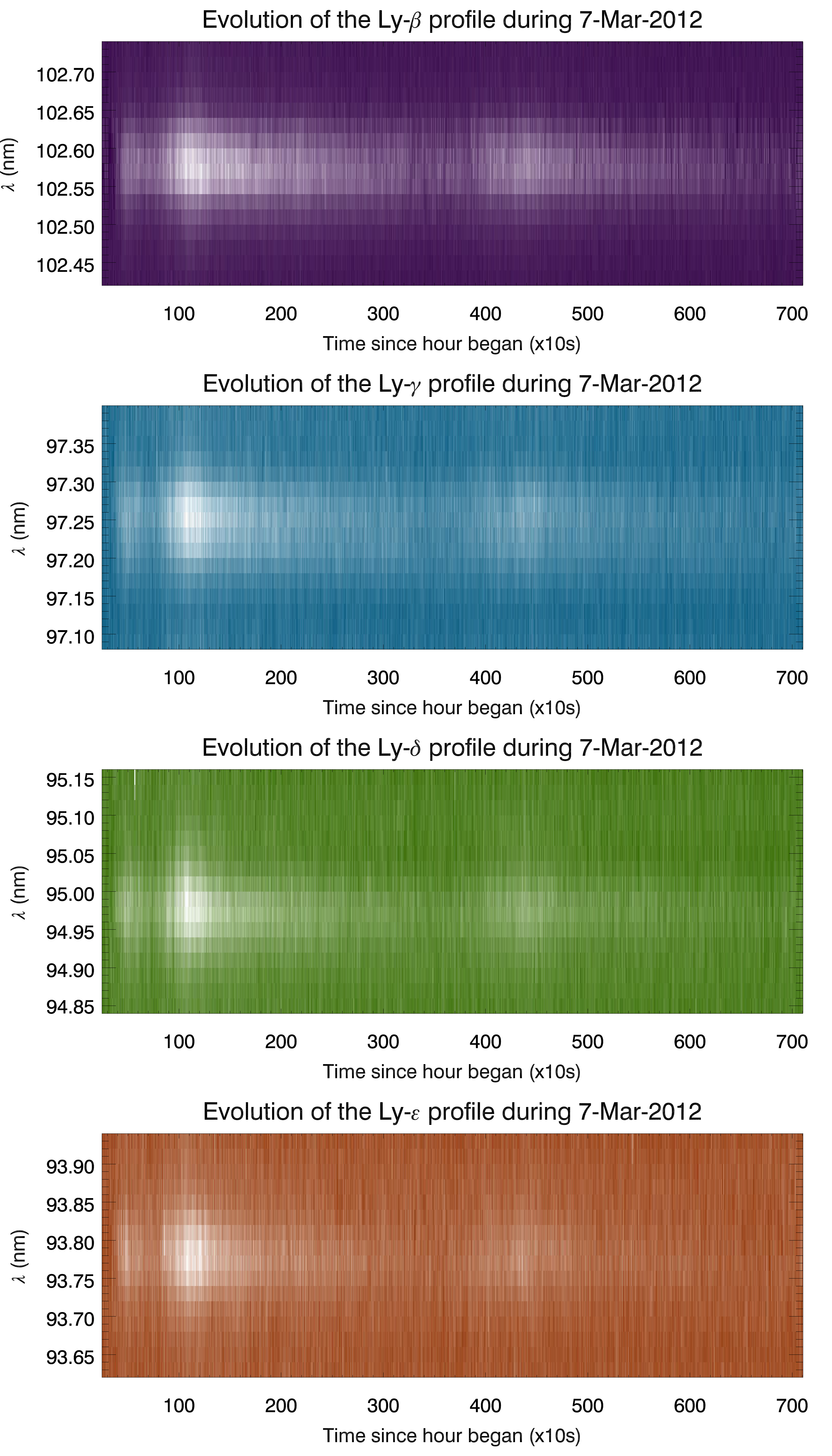}}
	\caption{Plots of the Lyman line irradiances in flare-excess for each wavelength bin at each time during the flare. Asymmetries can be seen at certain times, particularly in the profiles for Ly-$\delta$ and Ly-$\epsilon$.}
	\label{Fig4}
\end{figure}

While the majority of the line profiles are well fitted with a single Gaussian, they can exhibit asymmetries or bumps in their wings at certain times. This prompted a more thorough investigation of the line profile shapes. Figure \ref{Fig4} shows a  plot of the time-evolution of the normalised excess irradiance in each wavelength bin for each of the lines during SOL2012-03-07T00:07. We can see that there is variation in the irradiance of the wings as the flare progresses.

Clearly there is an asymmetry in some of the profiles such that not all of them are well-described by a single Gaussian at all times in the flare. From the CHIANTI line list (flare DEM) there were no expected strong lines in first or second order at these locations. We decided to attempt fitting with a double Gaussian to investigate the possibility of two plasma components at different speeds, emitting in the same line.

We assumed a double Gaussian line profile, with the Doppler shift of the stronger component constrained to less than 100 km s$^{-1}$ and its intensity greater than 70\% of the overall maximum intensity.   The weaker component was limited in intensity to between 10\% and 70\% of the overall maximum, with the lower bound chosen to avoid fitting noise. These constraints were arrived at by inspection of the data. 

In a minority of cases double Gaussian fits could be found, but no systematic behaviour was discovered. None of the flares exhibit any systematic or consistent evidence of a secondary ``moving" line component. The majority of secondary-component speeds either fluctuate often between upflows and downflows, or by inspection, are seen to be fits to noise. There is no clear evidence of a consistent high-speed component of the plasma motion. These time-varying asymmetries arise mainly due to noise. 

\subsection{Method 2 - Cross-correlation functions}

Our second method is based on matching patterns in the spectra, and does not make any assumptions about the shape of the line profiles. For two spectra, $f(\lambda)$ and $g(\lambda)$, their cross-correlation function is given in Equation \ref{eq1},

\begin{equation}
(f\star g)=\int_{-\infty}^{\infty}f^{\ast}(\lambda)g(\lambda+\Delta\lambda)d\lambda
\label{eq1}
\end{equation}

where the asterisk denotes the complex conjugate. The cross-correlation function (CCF) peaks when the lag ($\Delta\lambda$) is such that $g(\lambda+\Delta\lambda)$ most closely resembles $f(\lambda)$. If there were no Doppler shift, the CCF would peak at a lag of zero. Since we expect a Doppler shift, we also expect the CCF to peak at a non-zero lag.

We use the same flare intervals as in method 1, and carry out pre-flare subtraction to obtain flare excess in the same way. Then, for each time in the flare, the flare-excess Lyman line profiles are cross-correlated across their quiet-Sun counterparts over a range $\lambda-0.14 $ nm$ \leq \Delta\lambda \leq \lambda+0.14 $ nm.

\begin{figure}
	\resizebox{\hsize}{!}{\includegraphics{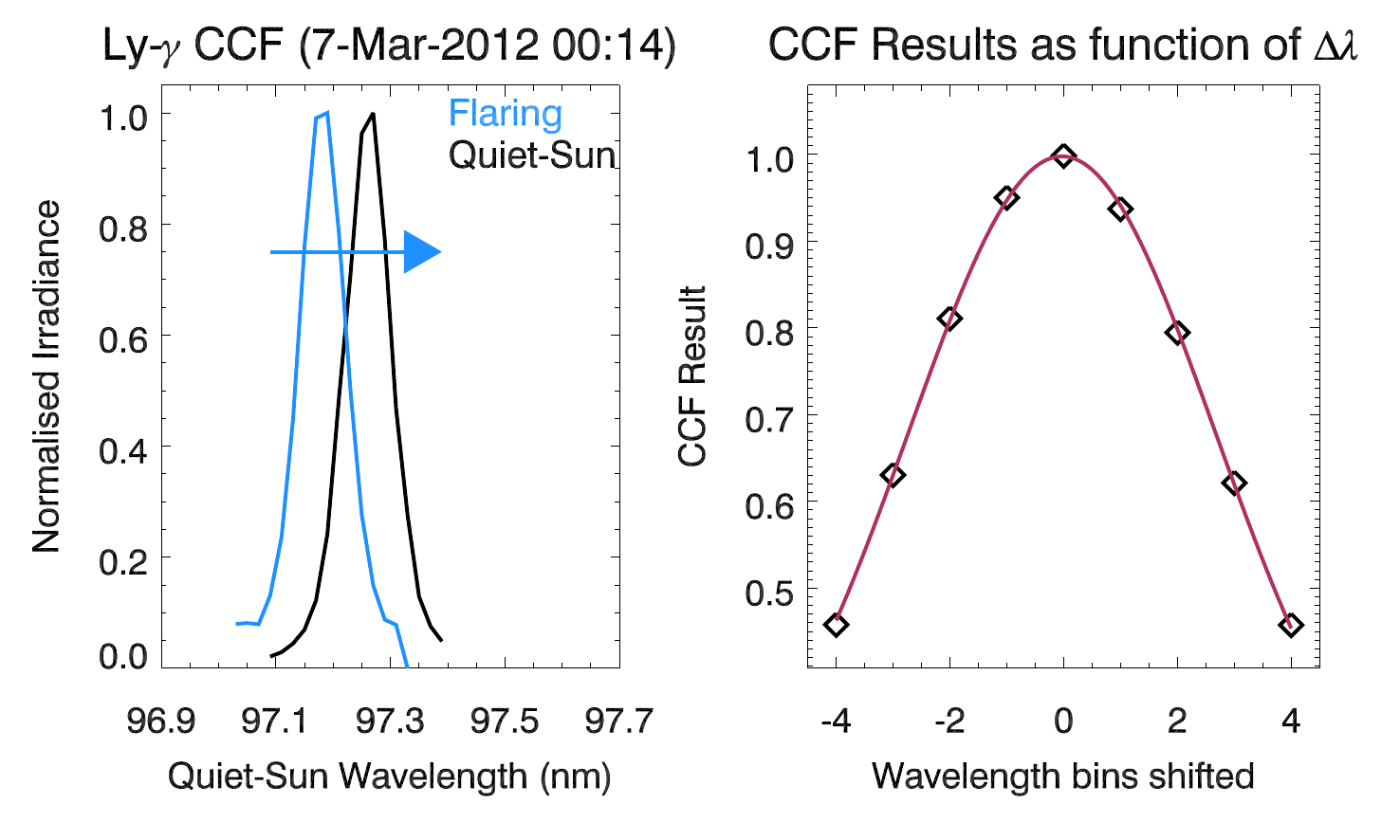}}
	\caption{Example of Cross-correlation process. The flare-excess Ly-$\gamma$ profile is moved across its Quiet-Sun counterpart, with the arrow indicating the direction of motion. The CCF results are fit with a Gaussian (plotted in maroon).}
	\label{Fig5}
\end{figure}

In order to shift the entire line profile by one wavelength bin size (0.02 nm), there would (via the Doppler formula) need to be a speed of around $65$ km s$^{-1}$, relatively high for the chromosphere. This would perhaps imply that all the CCF peaks would be recorded at a lag of 0 - not because there is no shift, but because it is below the resolution of the EVE data. However, it is possible to find the peak in the cross-correlation function at the sub-pixel level. We fit a Gaussian to the CCF results 4-points either side of the peak result. The value of the lag at which the Gaussian peaks (given by its centroid), which usually occurs between wavelength bin positions, is then taken as the Doppler shift, and is converted into a speed via the Doppler formula. The process is illustrated in Figure \ref{Fig5}.

In order to calculate the formal errors on the speeds for method 2, the procedure of \citet{Peterson1998} is used, in which uncertainties on the cross-correlation lags are estimated using a statistical realization process. Synthetic spectra are realized by altering the original spectrum by irradiance deviations based on the irradiance error described earlier. These realizations are correlated and a spread of centroid values is obtained. This is then used to calculate the error in the centroid on the actual data.

\subsection{Method 3 - Intensity-weighted centroids}

\begin{figure}
	\resizebox{0.85\hsize}{!}{\includegraphics{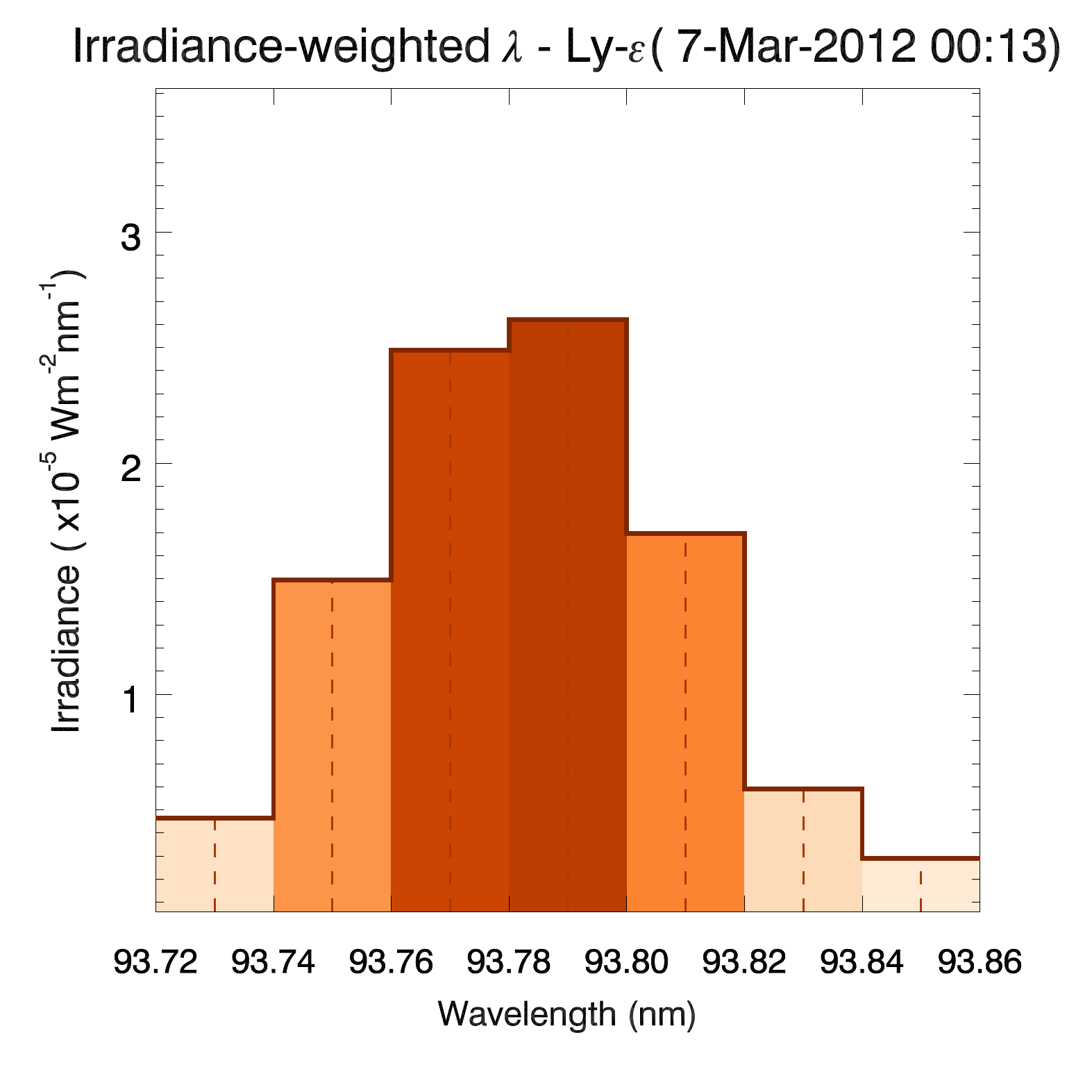}}
	\caption{An example of the weightings applied to each wavelength bin in order to calculate the centroid for the Ly-$\epsilon$ profile during the flare.}
	\label{Fig6}
\end{figure}

The third method uses the ``center of mass" of the line profiles as a means to measure their centroids. The positions of each wavelength bin are weighted by the irradiance in each bin, as illustrated in Figure \ref{Fig6}. The mean wavelength of the line profile is then calculated via Equation \ref{eq2}.

\begin{equation}
\bar{\lambda}=\frac{\Sigma_{i}I(\lambda_{i})\lambda_{i}}{\Sigma_{i}I(\lambda_{i})}
\label{eq2}
\end{equation} 

In the ``flare-excess" approach, the pre-flare irradiance values are again subtracted from each of the flaring spectra.  During times of low-irradiance enhancements, some bins contain a negative irradiance in ``flare-excess" and these are not included in the calculation as they result in spurious $\bar{\lambda}$ values. 

 $\bar{\lambda}$ is first calculated for each of the Lyman line profiles in the pre-flare averaged spectrum. For each time during the flaring period, each of the Lyman line profiles have their flaring $\bar{\lambda}$ calculated and the pre-flare $\bar{\lambda}$ subtracted. This gives us the Doppler shift, which is again converted into a speed using the Doppler formula.

This method appears to obtain lower speed results than the other two. This is likely because the intensity-weighting  method is by its nature more influenced by large intensity enhancements in the core of the line, with respect to the wings, which could lead to smaller calculated centroids.  It is therefore likely to provide a lower limit to the speed. 

\subsection{A comment on errors and uncertainties}
Formal errors are calculated by propagating the standard irradiance error through Equation \ref{eq2}, using the standard method for error propagation. It should be noted at this point that although each of the described methods allows calculation of a formal error on the Doppler speed, they should be considered with caution. Simple calculations of the variations in the line centroid positions during quiet-Sun conditions show larger speed fluctuations than are found using formal error propagation. 10 MEGS-B spectra were used to calculate the line-centroid positions over quiet-Sun conditions using each of the three methods, and the standard deviations of these positions were used to calculate conservative estimates of the error in Doppler speed for each of the lines. These errors are usually larger than the formal errors that can be calculated for each method by propagating the noise in the spectral irradiance through the various procedures used. 

This tells us that the errors on our speed measurements will in fact be dominated by the intrinsic spread in line-centroid positions. This spread is due to random motions on the solar surface in addition to instrumental and photon-counting uncertainties. An overview of the errors in quiet-Sun conditions and those calculated via the formal error propagation can be found in Table 2. 

 It is also known that the daily orbital motion of the spacecraft introduces  a variation in the  measured speeds of 3 km s$^{-1}$ \citep{HudsonEVE}. However, over the one-hour duration of one MEGS-B data file, this  variation  will be less than 1 km s$^{-1}$, and even less significant over the relatively short timescale of a flare. There could in principle also be wavelength variations caused by thermal or mechanical fluctuations within the instrument, however again we would expect these to be small over the short timescale represented by a flare, and not correlated with flaring activity (as the total increase in thermal load on EVE due to a flare will be tiny).  As our line shifts are calculated with respect to the pre-flare line centroid, and not with respect to an  absolute reference wavelength value, the speed during an event, relative to the pre-flare mean Sun-as-a-star speed, will be reliable to within about 1 km s$^{-1}$

\section{Results and Discussion}
We show the results for the 6 flares  selected as described in Section~\ref{section:methods}.
We have measured Doppler speeds using all three methods and both ``Sun-as-a-star" and ``flare-excess" data using the Lyman lines and the \ion{C}{iii} line in these events. \ion{C}{iii} is used for comparison as it is a very strong line and measurements can be made using the same three methods with lower errors (indicating also that our Lyman line measurements are noise-limited). 

\begin{figure*}
	\vbox{
		\hbox{
			\subfigure[]{ \resizebox{0.5\hsize}{!}{\includegraphics{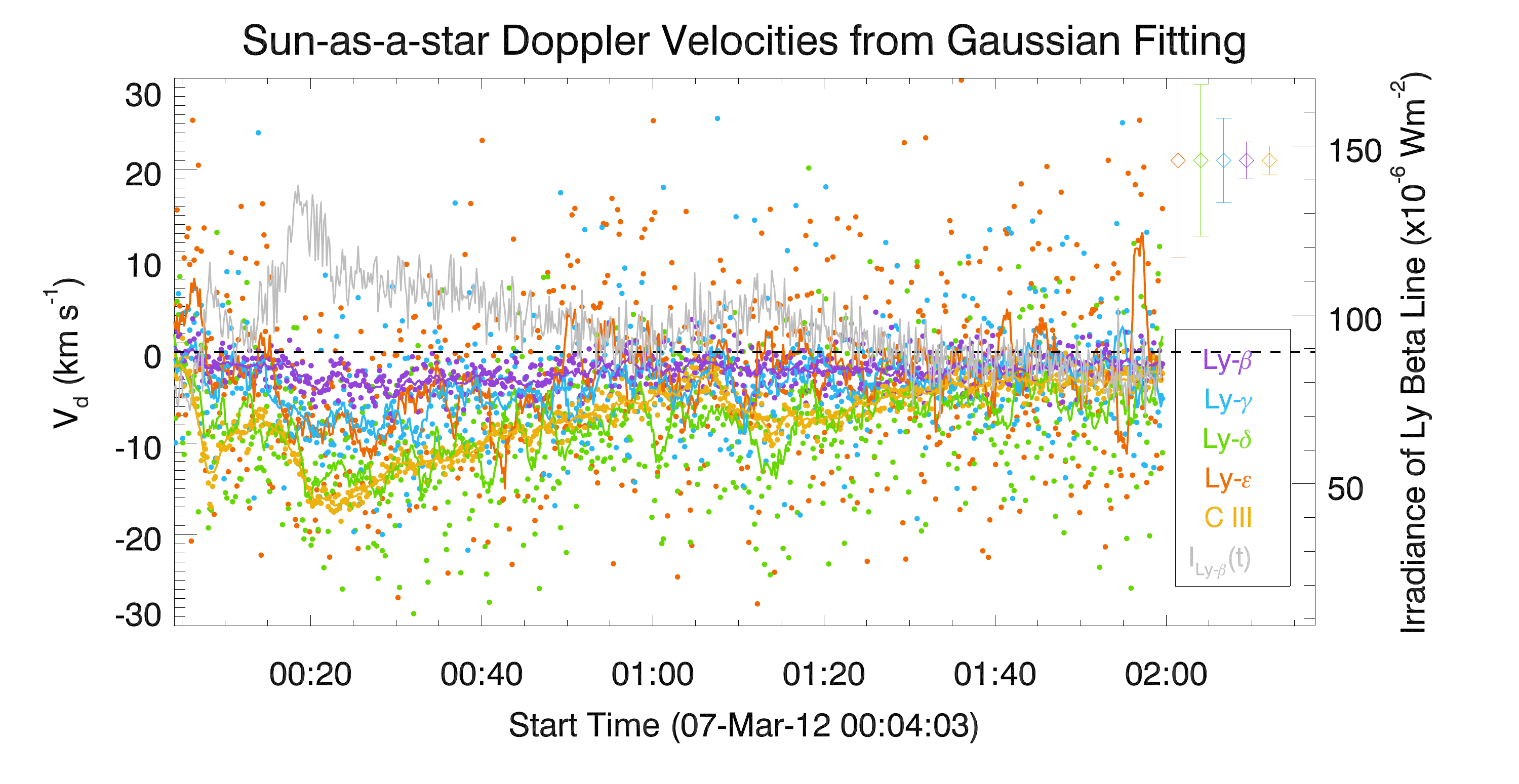}}  }  \subfigure[]{\resizebox{0.5\hsize}{!}{\includegraphics{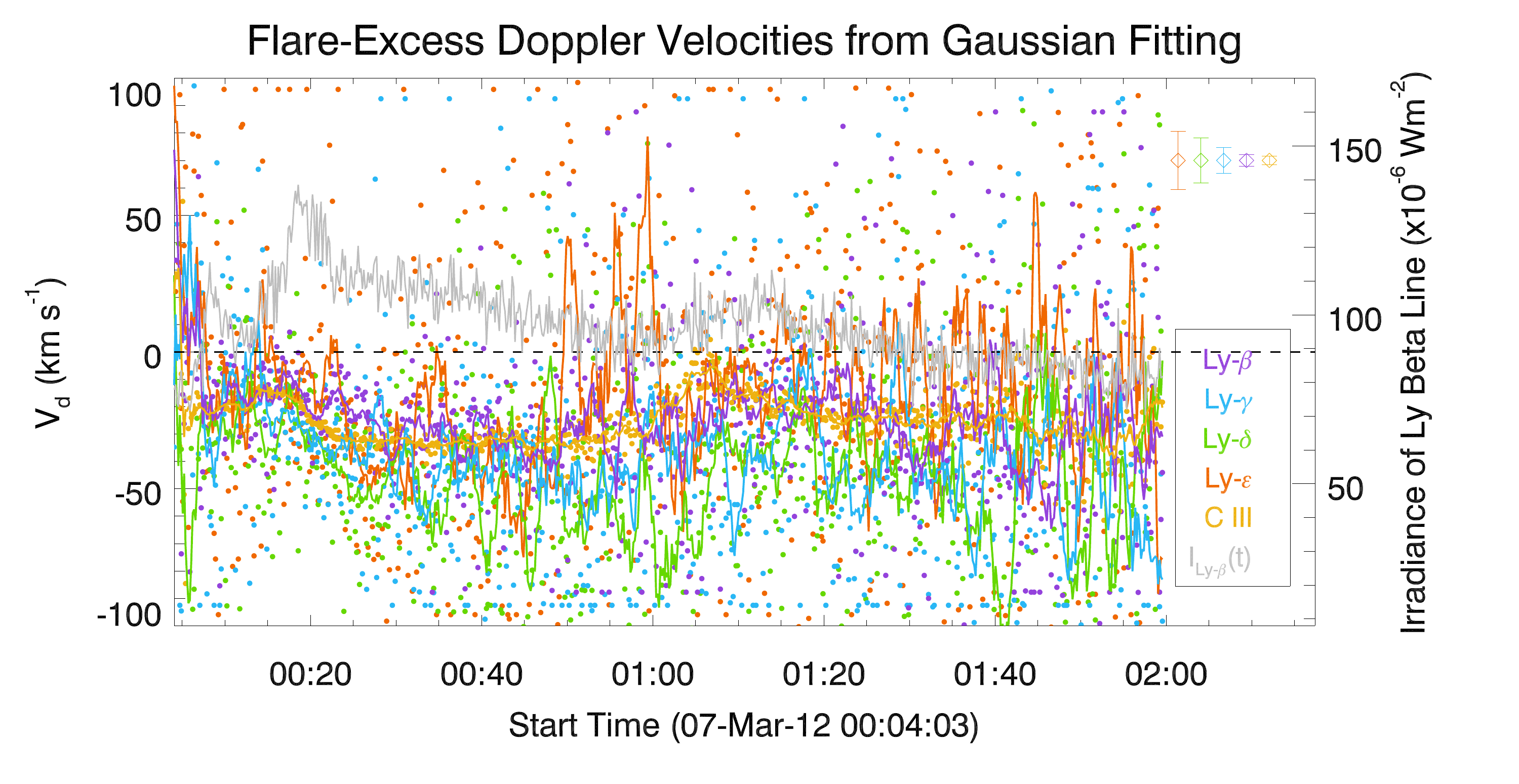}}  }} 
		\hbox{
			\subfigure[]{\resizebox{0.5\hsize}{!}{\includegraphics{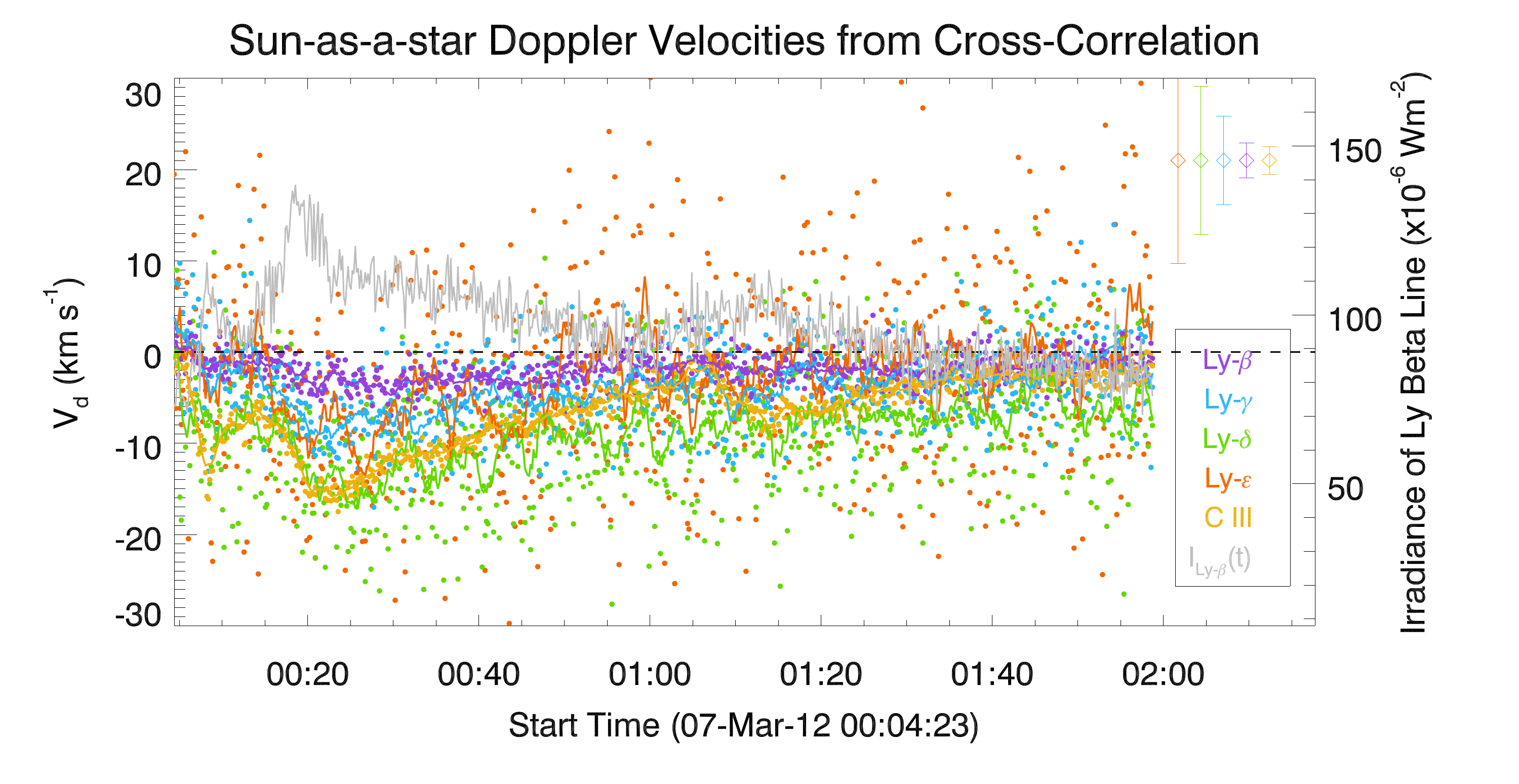}} } \subfigure[]{\resizebox{0.5\hsize}{!}{\includegraphics{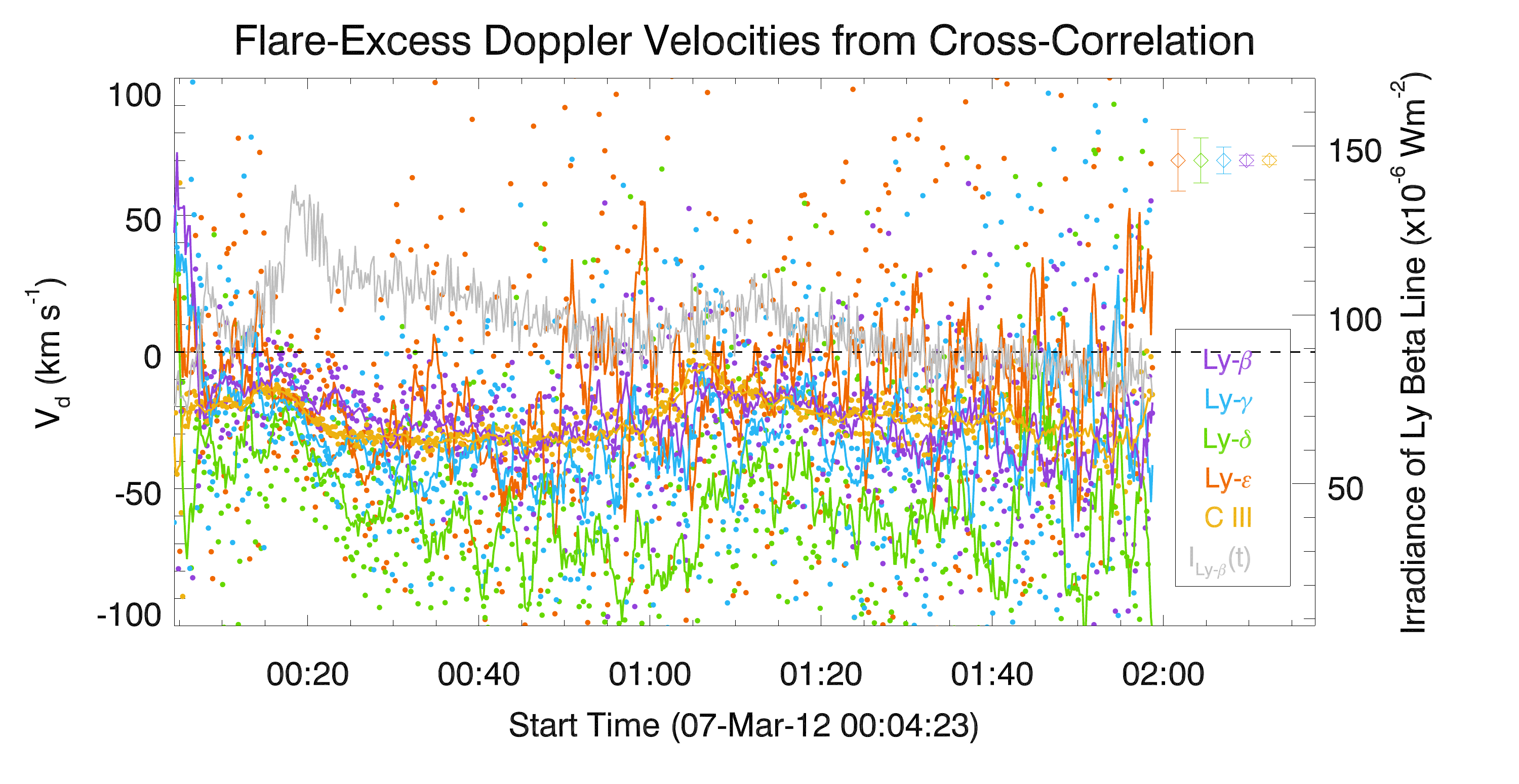}} }} 
		\hbox{
			\subfigure[]{\resizebox{0.5\hsize}{!}{\includegraphics{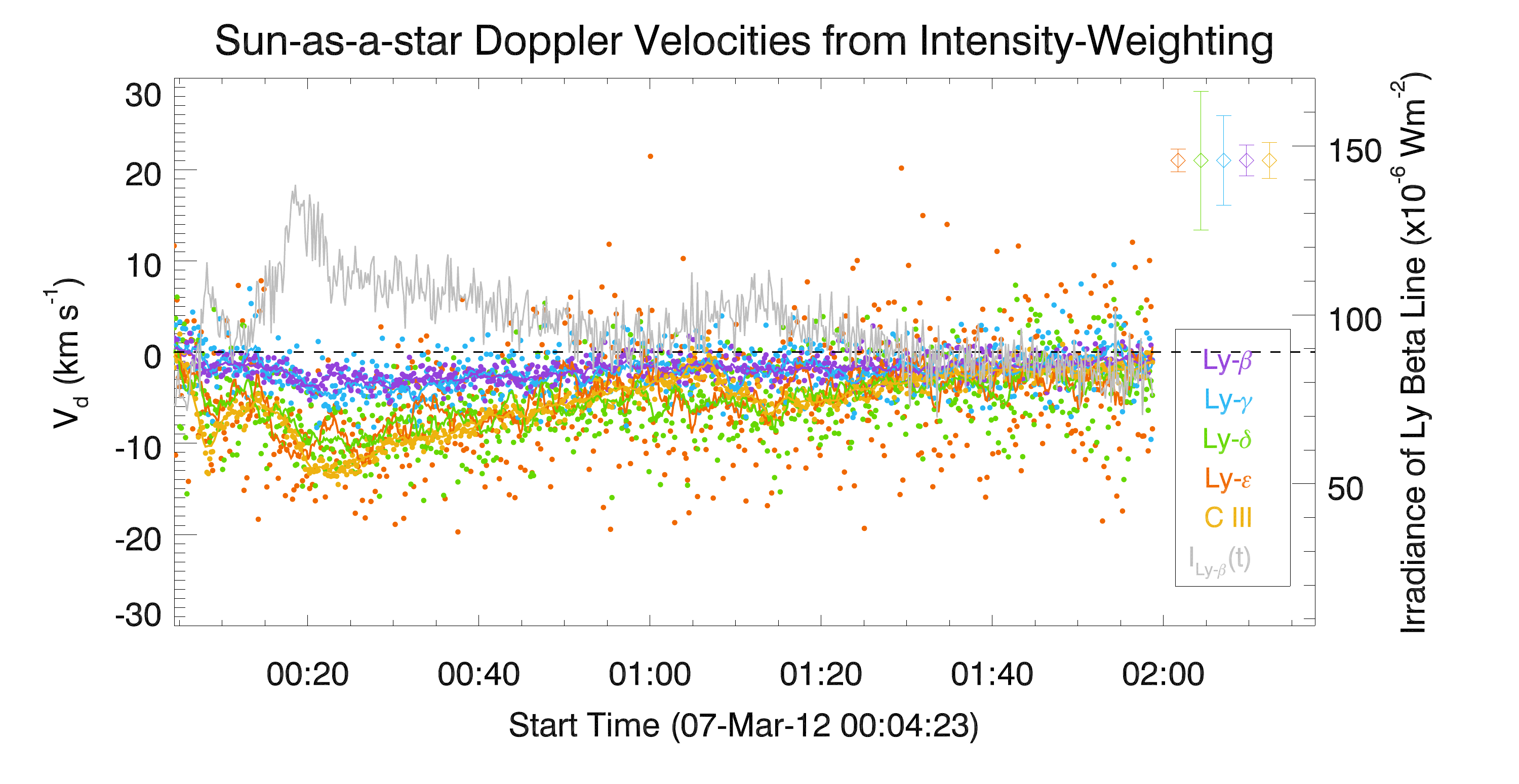}} } \subfigure[]{\resizebox{0.5\hsize}{!}{\includegraphics{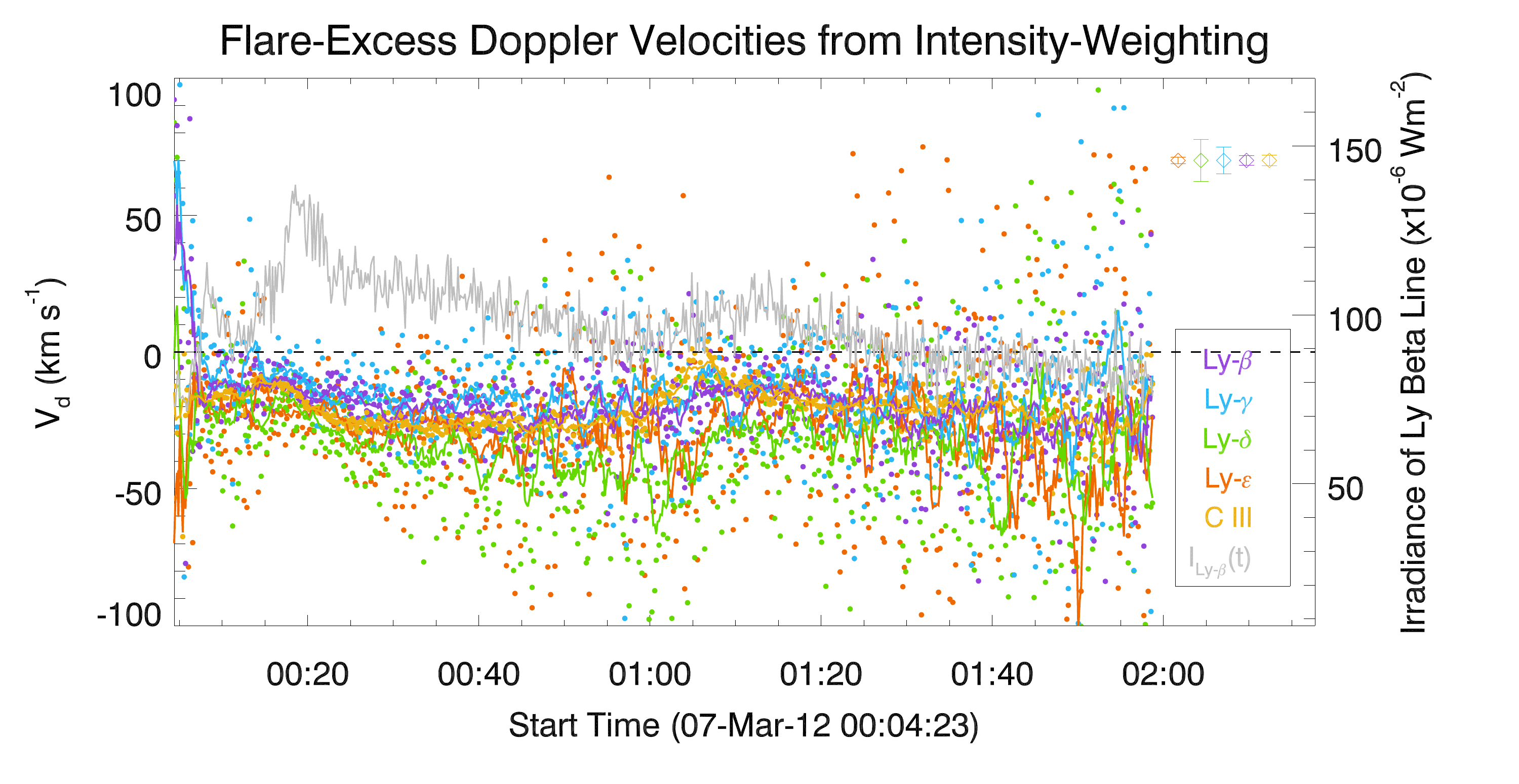}} }} 
	\caption{Doppler speeds for the SOL2012-03-07T00:07 X5.4 flare. The left hand column displays results for ``Sun-as-a-star" data, while the right hand column exhibits results in ``Flare-excess". From the top row to the bottom, results are obtained for methods 1, 2 and 3 respectively. Representative error bars are plotted in the upper right of the figures, which display the size of the average error for each data-series. Velocities are smoothed with a boxcar of 10. Negative speeds are blueshifts.}
		\label{Fig7}
	}
\end{figure*}

We provide results for both ``Sun-as-a-star" and ``flare-excess" data'' for the X5.4 SOL2012-03-07T00:07 flare in Figure \ref{Fig7}. Representative error bars are plotted in the top-right corner of the graphs. These show the average size of the errors on the speeds for each of the measured lines. The errors used here are those derived from the variations in quiet-Sun line centroid positions and not those calculated from the formal fitting procedures.  This is because the intrinsic solar variations are dominant error terms. 

\begin{figure*}
			\subfigure[]{\resizebox{\hsize}{!}{\includegraphics[]{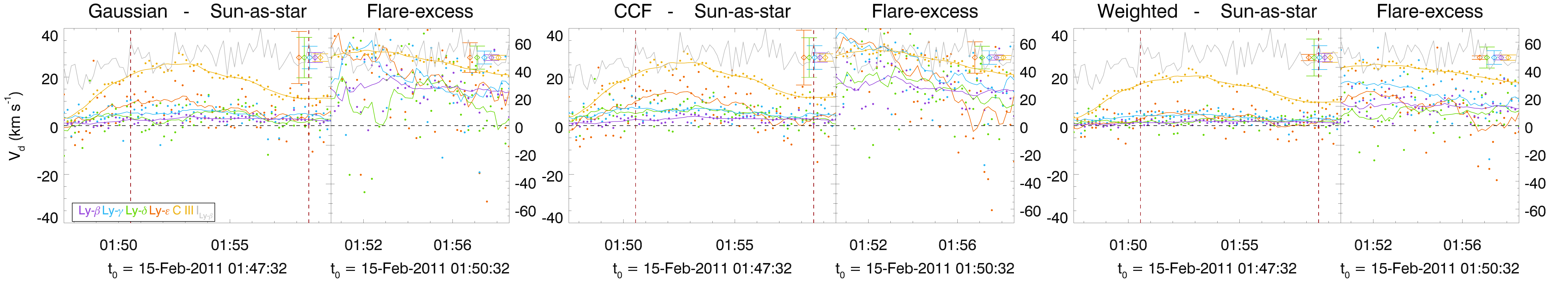}}}  \subfigure[]{\resizebox{\hsize}{!}{\includegraphics[]{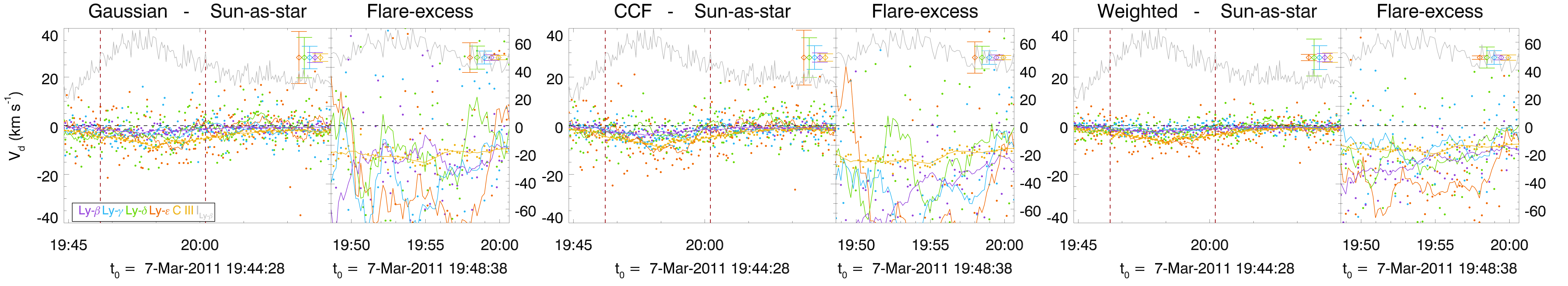}}}
			\subfigure[]{\resizebox{\hsize}{!}{\includegraphics[]{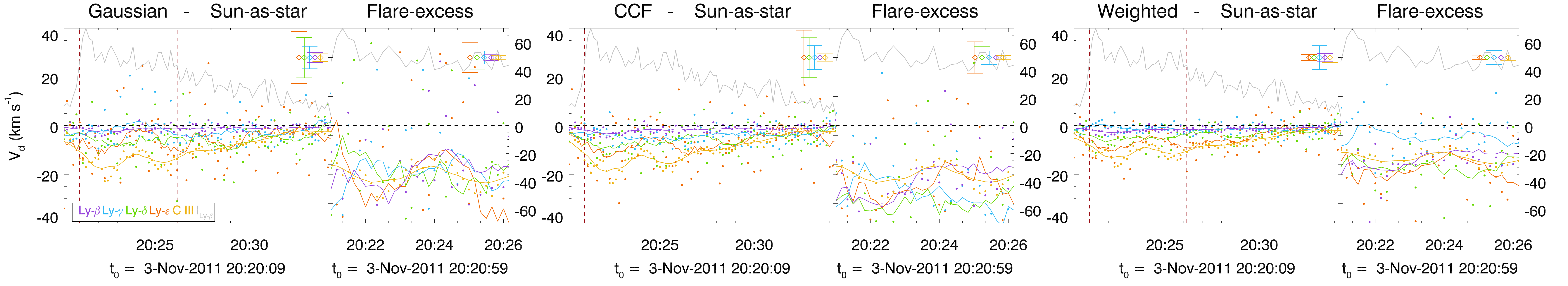}}}  
			\subfigure[]{\resizebox{\hsize}{!}{\includegraphics[]{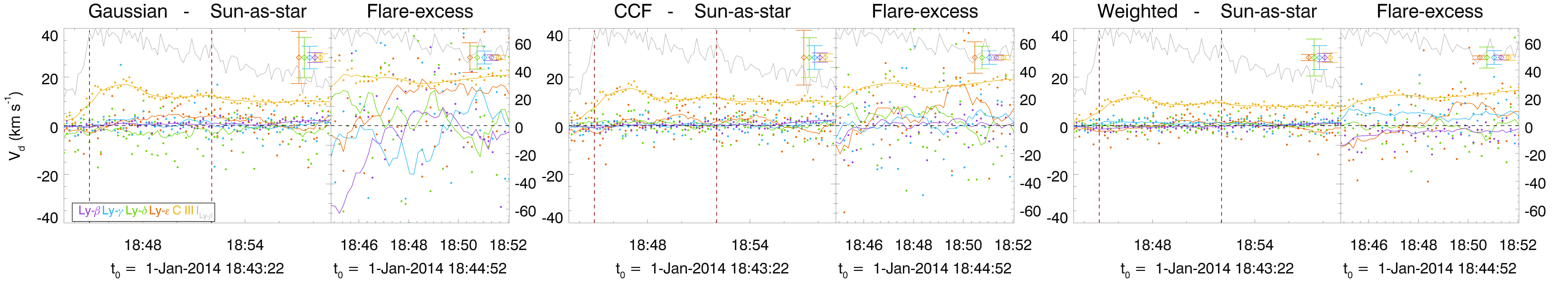}}}
			\subfigure[]{\resizebox{\hsize}{!}{\includegraphics[]{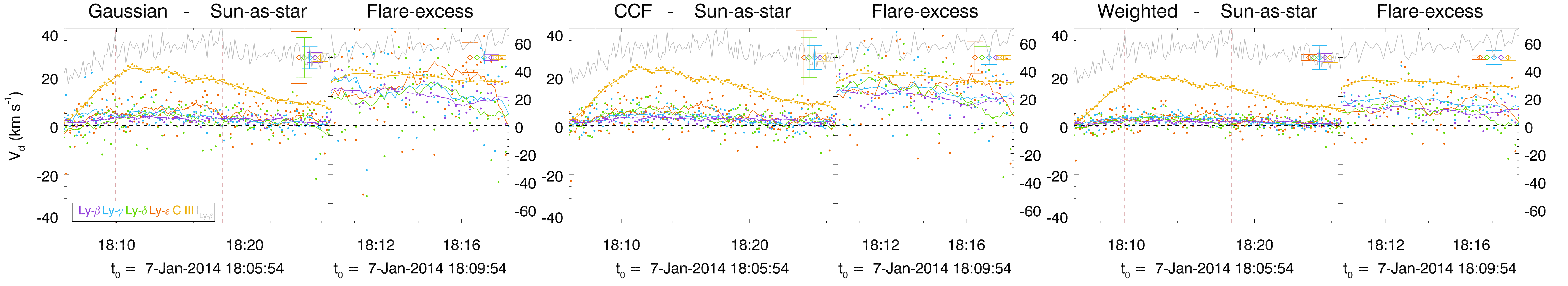}}}
			
	\caption{Doppler speeds for the remaining 5 flares studied. Each subfigure shows speeds calculated for each method. ``Sun-as-a-star" speeds are plotted in the left-hand panels, while ``flare-excess" speeds are on the right. The ``flare-excess" speeds are plotted for only a short period of time, corresponding to when the lightcurve in Ly-$\beta$ is $2\sigma$ above its pre-flare mean. The ``flare-excess" time-ranges are indicated on the ``Sun-as-a-star" plots by the dashed vertical lines. Overplotted in grey are the normalised Ly-$\beta$ lightcurves.}
	\label{Fig8}
\end{figure*}

 Of the 6 flares studied, the X5.4 flare SOL2012-03-07T00:07  (Figure \ref{Fig7}) displays the clearest velocity signal of each of its spectral lines across all three methods and in both ``Sun-as-a-star" and ``flare-excess" signals. We observe clean velocity profiles for the lines in the ``Sun-as-a-star" data, typically of the order 10 km s$^{-1}$ Figure \ref{Fig7}, left-hand column), while the velocity profiles in the ``flare-excess" graphs (Figure \ref{Fig7}, right-hand column) have a higher magnitude due to subtraction of the full-disk contributions which are stationary on average, but also a greater variability and noise due to a lower signal overall being fitted. However, until around 01:30 the same underlying patterns can be seen as in the ``Sun-as-a-star" graphs. The large and fluctuating ``flare-excess" speeds close to the start and end of the event are not real, but result from fitting the very noisy line profiles obtained when the pre-flare line profile is subtracted from a signal which is not yet enhanced, or is decaying in intensity back towards its pre-flare level. This can be seen clearly in Figure \ref{Fig7}f, where the increased variability is coincident with the times where the irradiance is low. After around 01:30, the Lyman line intensities have returned to their pre-flare levels and the ``flare-excess"  speeds are clearly dominated by noise.

This flare displays strong blueshifts in all of the lines considered, although intensity-weighted centroiding sometimes tends to produce lower speeds compared to those found  using the other two methods. Considering the ``flare-excess" graphs (and hence the true velocity signals), upflows in Ly-$\beta$ of around 20-30 km s$^{-1}$ are observed. Ly-$\gamma$ also exhibits upflows of around 30-40 km s$^{-1}$, albeit with greater variability (intensity-weighted centroiding  finds around 20 km s$^{-1}$  in Ly-$\gamma$). Ly-$\delta$  speeds are highest, increasing from around 30 km s$^{-1}$ at flare onset to 50 km s$^{-1}$ at its peak, although it does display significant variability in its speed. Ly-$\epsilon$ also appears highly variable (likely due to its location in a relatively crowded region of the spectrum, which makes fitting difficult) but tends to average between 10-20 km s$^{-1}$. The \ion{C}{iii} line exhibits consistent behaviour across all three methods, showing upflows of around 20 km s$^{-1}$ at the start of the flare, which peak at around 30 km s$^{-1}$. 
 
  For the remaining 5 flares, in Figure \ref{Fig8} we show the ``Sun-as -a-star" speeds for the full duration of the flare and the ``flare-excess" for a shorter time-period, corresponding to when the Ly-$\beta$ is more than 2-$\sigma$ above the background and a value can be found reliable. In Figure \ref{Fig8}, each row is a different flare, and each column a different method. A summary of the mean ``flare-excess" speeds,  obtained using method 2 and averaged across this restricted time period, is given in Table \ref{Table1}.  
 
{\bf SOL2011-02-15T01:45}: Prominent redshifts are observed in all lines shown during this flare (Figure \ref{Fig8}a).  The ``Sun-as-a-star" downflows peak at 3 km s$^{-1}$ in Ly-$\beta$, 6 km s$^{-1}$ in Ly-$\gamma$ and Ly-$\delta$, and 12 km s$^{-1}$ in Ly-$\epsilon$. The ``flare-excess" results suggest downflows of a greater magnitude, typically of about 30-50 km s$^{-1}$. \ion{C}{iii} displays the greatest speeds during this flare, attaining 26 km s$^{-1}$ in ``Sun-as-a-star" and around 50 km s$^{-1}$ in ``flare-excess".
 
 {\bf SOL2011-03-07T19:46}: Clear blueshifts are observed during this flare (Figure 8b), with the speed obtained from Ly-$\beta$ increasing to 4 km s$^{-1}$ in the ``Sun-as-a-star" data, then decaying. Similar rise-decay behaviour in the speed is generally observed in all of the lines in this flare, with increasing maximum speeds for higher order lines. The \ion{C}{iii} line displays a stable rise-decay profile with a maximum velocity of 9 km s$^{-1}$ during this flare. When the ``flare-excess" data is considered for this flare, blueshifts are again observed in all lines with a similar rise-decay behaviour, but with typical maximum speeds of 60-70 km s$^{-1}$. The only exception in ``flare-excess" is \ion{C}{iii}, which displays a less-variable acceleration to 26 km s$^{-1}$.
 
{\bf SOL2011-11-03T20:20}: The Lyman lines again display systematic upflows (Figure 8c), with observable rise-decay behaviour observed in their velocity in the ``Sun-as-a-star" data. Ly-$\beta$ displays the slowest speeds in this flare, averaging around 2 km s$^{-1}$. The higher-order lines have greater maximum speeds,  with Ly-$\gamma$, $\delta$ and $\epsilon$ attaining 6, 10 and 12 km s$^{-1}$ respectively. When ``flare-excess" is considered, the true speeds appear to be on the order of 30-50 km s$^{-1}$.  Ly-$\beta$ exhibits a similar speed profile across all three methods in ''flare-excess", with a peak speed of 50 km s$^{-1}$ in methods 1 and 2.  The fastest line in ``Sun-as-a-star" is \ion{C}{iii}, which accelerates to 19 km s$^{-1}$ before decaying. Additionally, this line appears to display periodic motion in ``flare-excess" about 35 km s$^{-1}$ with an initial amplitude of 10 km s$^{-1}$. 
 
  {\bf SOL2014-01-01T18:44}: Despite the Ly-$\beta$ lightcurve during this flare showing a particularly prominent irradiance enhancement the flows are very weak, or absent (Figure \ref{Fig8}d), but on average the Lyman lines appear to be redshifted. The speeds are generally below 5 km s$^{-1}$ in the ``Sun-as-a-star" data and around 10 km s$^{-1}$ in ``flare-excess". The ``flare-excess" speeds are variable, particularly those obtained using method 1. 
Observations of this event in the AIA $304$ $ \text{\AA}$ and $171$ $ \text{\AA}$ channels reveal a conspicuous ejection of material, which may have implications for the observed speeds. We discuss these in section $3.1$.

{\bf SOL2014-01-07T18:06} exhibits redshifts (Figure \ref{Fig8}e), with all of the Lyman lines showing clear downflow speeds of around 3-9 km s$^{-1}$ in ``Sun-as-a-star" data  and between 20-30 km s$^{-1}$ in ``flare-excess". This may be due to all of the lines emitting from the same volume of plasma. Speeds found from \ion{C}{iii} are  much higher than are found from the Lyman lines in both of the January 2014 flares, attaining ``Sun-as-a-star" speeds of 14 km s$^{-1}$ during SOL2014-01-01T18:44 and 23 km s$^{-1}$ during SOL2014-01-07T18:06. The ``flare-excess" data for \ion{C}{iii} suggests typical speeds of 20-30 km s$^{-1}$.

 \subsection{Search for flows and ejecta in AIA data and the literature}
  Speeds measured in these flares sometimes correspond to upflows, and sometimes to downflows along the line-of-sight. As discussed in the introduction, theoretical ideas about the response of the chromosphere to energy input suggest that the Lyman lines, being emitted by cooler plasma, should be redshifted on average because of the effect of the chromospheric `condensation'. However, material ejected outwards during a flare, e.g. in a filament eruption, could also contribute to the observed lineshifts.

\begin{figure*}
 	\subfigure[]{\vbox{  \hbox{   \resizebox{0.89\hsize}{!}{\includegraphics{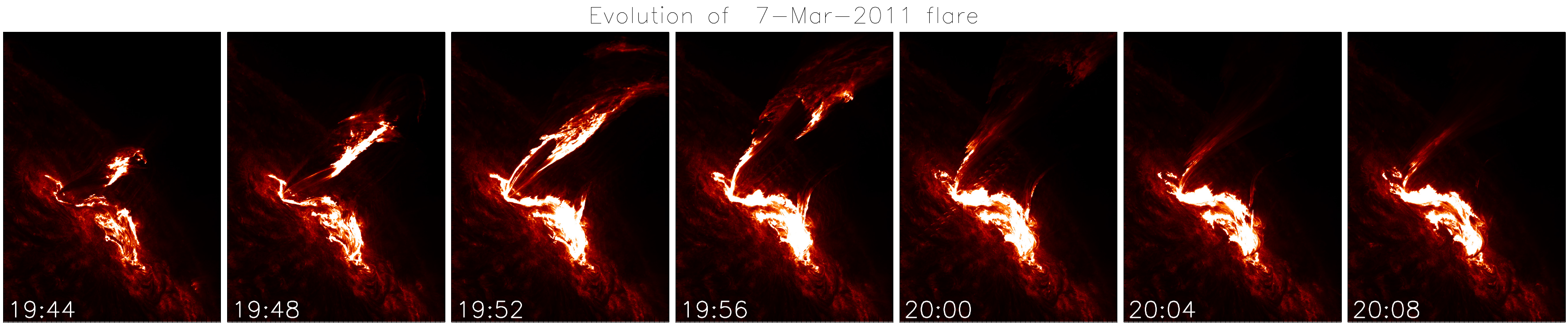}}  }} }
 	\subfigure[]{\vbox{  \hbox{   \resizebox{0.89\hsize}{!}{\includegraphics{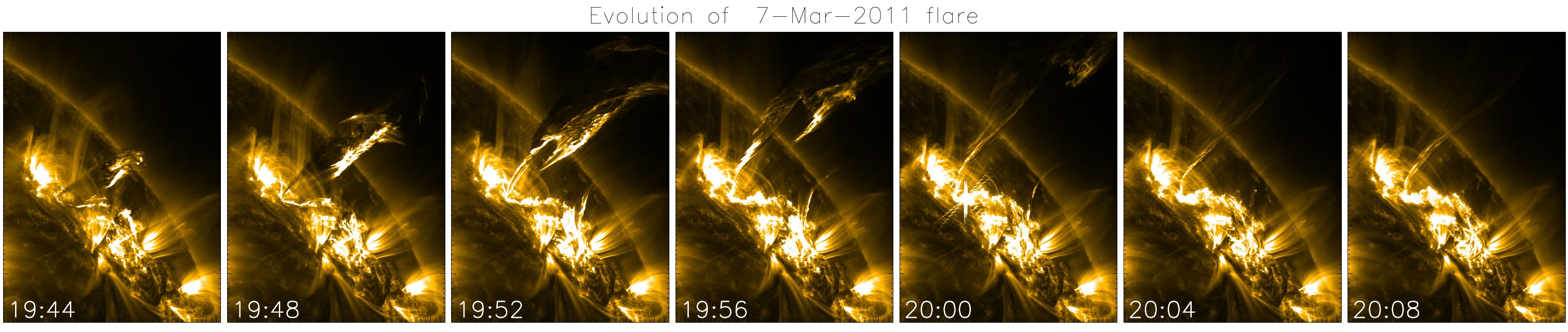}}  }} }
 	\subfigure[]{\vbox{  \hbox{   \resizebox{0.89\hsize}{!}{\includegraphics{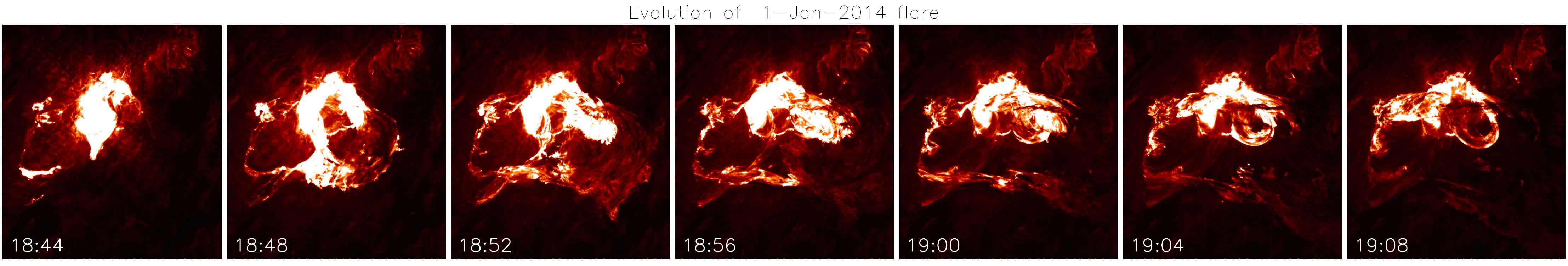}}  }} }
 	\subfigure[]{\vbox{  \hbox{   \resizebox{0.89\hsize}{!}{\includegraphics{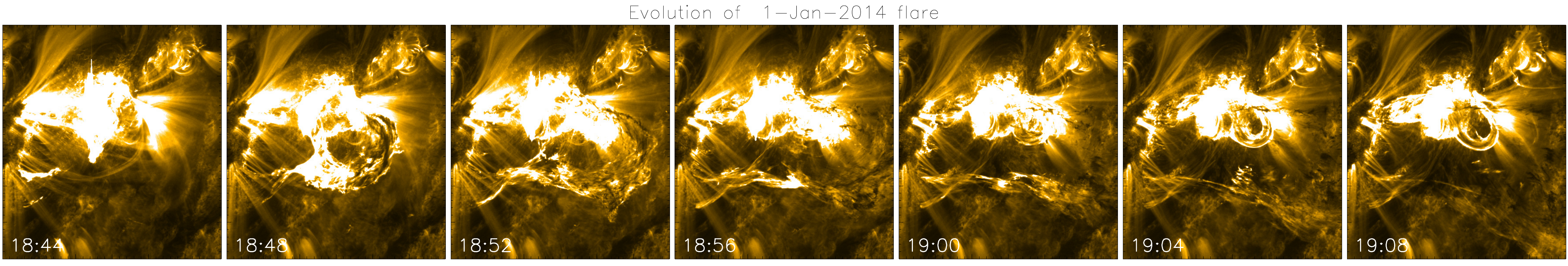}}  }} }

 	\caption{Notable flows during the 07th March 2011 and 01st January 2014 flares. These are the only two flares considered that display prominent ejections of material. Panels a and c show images in the 304 $\text{\AA}$ channel, while b and d show emission in the 171 $\text{\AA}$ channel.}
 	\label{Fig9}
 \end{figure*}

We use images of these events from the SDO's Atmospheric Imaging Assembly (AIA) instrument to examine the latter possibility.  We have inspected AIA movies in both $304$ $ \text{\AA}$ and $171$ $ \text{\AA}$, with the $304$ $ \text{\AA}$ passband showing motions of plasma at chromospheric temperatures, and $171$ $ \text{\AA}$ observing any motions of plasma in the hotter coronal magnetic loops and other heated ejecta.  In Figure \ref{Fig9} we present ``filmstrip" plots of notable features during  two of the six flares,  SOL2011-03-07T19:46 and  SOL2014-01-01T18:44 which may shed light on some of our results from EVE. In the other four flares, no notable flows, motions or ejecta are detected in movies or in difference movies. We have also searched the literature for reports of ejecta in these events.

{\bf SOL2011-03-07T19:46}:  This event, located at N30W48 shows upflows in the Lyman lines lasting about 15 minutes.  Shortly after onset, a large eruption of material leaves the active region, expanding and propagating upwards (Figures \ref{Fig9}a and \ref{Fig9}b).  This is clearly observed in both $171$ $ \text{\AA}$ and $304$ $ \text{\AA}$. The ejection of material is possibly a contributor to the upflow signatures observed in Figure \ref{Fig8}b. Tracking of the feature in the $304$ $ \text{\AA}$ channel yields a flow velocity of roughly $ v_{f}=250 $ km s$^{-1}$ in the plane of the sky, which is likely an underestimate due to the motion being projected.  This is large compared to the chromospheric velocities expected at low temperatures from the evaporation/condensation process. If this erupting plasma is moving roughly radially with a component of velocity directed towards the observer, and also emitting in the Lyman lines, they should be blueshifted. 

It is worth noting that the effects of Doppler dimming on resonant \ion{He}{i} and \ion{He}{ii} in outward-moving prominence material were investigated by \citet{Labrosse2007}, who observed both decreased overall intensity in these lines accompanied by a red-wing enhancement. This effect will also be present in the H Lyman lines \citep{Heinzel1987, Gontikakis1997}. Similarly, any surface-directed plasma may be subject to Doppler dimming, resulting in an enhancement in the blue wing. However, it is unlikely that there are any significant bulk motions towards the surface in this event. Finding an explanation for the blueshifted emission in this flare remains challenging.

{\bf SOL2014-01-01T18:44} This flare, located at S16W45, displays the weakest flows in the Lyman lines with overall a small tendency for downflows (Figure \ref{Fig8}d).  AIA images show a prominent ejection of material towards the west in both $171$ $ \text{\AA}$ and $304$ $ \text{\AA}$, along with material flowing along loops (Figures \ref{Fig9}c and \ref{Fig9}d). Given the clear ejection of material, it is curious that we do not observe strong upflows in the Lyman or \ion{C}{iii} lines. The weak flows in the Lyman lines could be due to the net result of downflowing chromospheric plasma producing strong emission, and weak emission from upflowing ejected plasma, as is observed in the AIA images.  As with SOL2011-03-07T19:46, we tracked the eruptive feature and found a flow velocity in the plane of the sky of roughly $v_{f}=130$ km s$^{-1}$. This is again a high velocity relative to expected chromospheric motions, and it is possible that eruptive material emitting in the Lyman lines is contributing a blueshifted component to the line profiles.

{\bf SOL2011-02-15T01:45} is located at S20W10 and shows redshifts in all lines examined, consistent with downflows. In this much-studied event, the $171$ $ \text{\AA}$ channel reveals loop ``contractions" both north and south of the active region shortly after flare onset,  as shown for example in \citet{2012ApJ...749...85G} and \citet{2012ApJ...748...77S}. These motions of 20-40~km~s$^{-1}$ towards the core of the active region last about 5 minutes. Conceivably, the downflows observed in the Lyman lines during this flare (Figure \ref{Fig8}a) are also associated with this process. Non-linear force-free extrapolations around the event by \citet{2012ApJ...748...77S} show evidence for a magnetic flux rope, the altitude of which decreases around the flare impulsive phase, but \cite{2011ApJ...738..167S} comment that there is no sign of prominence material in AIA 304\text{\AA}. The flare is associated with a large CME.

{\bf SOL2011-11-03T20:20} at N21E64 displays strong blueshifts in Figure \ref{Fig8}c. \cite{2013ApJ...778...70C} and \cite{2014ApJ...790....8L} identify this event as a confined flare and failed filament eruption using stereoscopic data.  \cite{2013ApJ...778...70C} shows that between 20:20UT and 20:25UT a very small heated filament, which is bright in 304~\text{\AA}, becomes unstable and moves roughly to the north-east at projected speeds of up to 400~km~s$^{-1}$.  If it had a small velocity component in the line of sight this could account for some of the flows we measure, though they continue for longer. 

{\bf SOL2012-03-07T00:07} located at N30W48 has very strong blueshifts (Figure \ref{Fig7}). This was a major eruptive event with a CME and Type II radio burst \citep{2014JGRA..119.6042S}.  Observations in the AIA $171$ $ \text{\AA}$ channel reveal a large amount of loop motion during this flare, with an appreciable loop expansion east and west of the active region. The reported eruption  and the loop motions may contribute to the observed upflows.

{\bf SOL2014-01-07T18:06} at S12W08, exhibits significant Lyman line redshifts.  This event was the source of a rapid CME, estimated to be directed between 30$^{\circ}$ and 50$^{\circ}$ from the Sun-Earth line \citep{2015ApJ...812..145M}, so we would expect this to be associated with strong upflows in the ejecta. Our AIA observations of this flare show very little motion in the active region. We detect  ribbon brightenings, with a very faint downflow visible in the difference images, which may be consistent with the downflows detected by EVE.

 \section{Conclusions}
 
 \begin{table*}
 	\centering
 	\begin{tabular}{c c c c c c c c}	
 		\hline \hline
 		Flare Date/Time & GOES Class &  Location & $\bar{V_{\beta}}$ & $\bar{V_{\gamma}}$ & $\bar{V_{\delta}}$
 		& $\bar{V_{\epsilon}}$ & $\bar{V_{\ion{C}{iii}}}$  \\
 		\hline
 		SOL2011-02-15T01:45 & X2.2 & S20W10 &  27  & 43   & 26  & 35   & 48 \\
 		SOL2011-03-07T19:46 & M3.7 & N30W48 & -44 & -44  & -26 & -50 & -23 \\
 		SOL2011-11-03T20:20 & X1.9 & N21E64 & -39 & -45 & -54   & -42  & -36 \\
 		SOL2012-03-07T00:07 & X5.4 & N18E31 & -17 & -26  & -32 & -18  & -21 \\
 		SOL2014-01-01T18:44 & M9.9 & S16W45 & 1   & 2  & 8   & 8  & 29 \\   
 		SOL2014-01-07T18:06 & X1.2 & S12W08 & 21  & 24   & 23  & 27   & 35 \\
 	\end{tabular}
\caption{Time-averaged Doppler Velocities (in km s$^{-1}$) calculated using Method 2 on ``flare-excess" spectra for each of the 6 flares considered. The speeds have be time-averaged over the period during which the Ly-$\beta$ lightcurve is $2\sigma$ above the pre-flare average.}
\label{Table1}
 \end{table*}
 
 \begin{table*}
 \centering
    \begin{tabular}{c c | c c c c c}
 	\hline \hline
 	Method &  & Ly-$\beta$ & Ly-$\gamma$ & Ly-$\delta$ & Ly-$\epsilon$ & \ion{C}{iii} \\
 	\hline
 	{\multirow{2}{*}{1}} & Quiet-Sun & 2.02 & 4.64 & 8.30 & 10.69 & 1.57 \\
 	                       & Formal    & 1.06 & 3.84 & 5.58 & 7.95 & 0.13 \\ \hline
 	{\multirow{2}{*}{2}} & Quiet-Sun & 1.90 & 4.88  & 8.14  & 11.28  & 1.52 \\
 	                        & Formal    & 1.40 & 2.69 & 3.17 & 3.97 & 0.73 \\ \hline
 	{\multirow{2}{*}{3}} & Quiet-Sun & 1.70 & 4.90 & 7.60 & 1.27 & 1.95 \\
                         	& Formal    & 0.03 & 0.10 & 0.25 & 0.56 & 0.01 \\ \hline
 	\end{tabular}
	\caption{Errors for each method have been derived from analysis of the spread of line-centroids during quiet-Sun conditions. Time-averaged formal errors, averaged over flares studied for each of the three methods are also stated.}
	\label{Table2}
  \end{table*}

 We have used three independent methods to examine the Doppler shifts and plasma speeds in the hydrogen Lyman and the \ion{C}{iii} lines during 6 M and X class flares. These lines are typically chromospheric \citep{Milligan2015} and observations of systematic flows in these lines could  be related to the energy deposition and heating in this part of the solar atmosphere.  Alternatively, they may be related to mass motion in the line-of-sight of chromospheric-temperature plasma, in an ejection or filament motion, not easily measurable by other means.

We tabulate time-averaged ``flare-excess" speeds for each of the lines calculated using method 2 in Table \ref{Table1}, where the time-averaging is done for a duration of time corresponding to the Ly-$\beta$ lightcurve being $2\sigma$ above its pre-flare average.  From this we see that 3 of the flares mainly exhibit upflows, and the other 3 exhibit downflows. We also note that the \ion{C}{iii} line also moves in the same direction as the Lyman lines in each of the flares. 
 
Although we have been able to detect systematic flows in each of the lines for all flares studied, we do not observe any strong tendency for flows to be predominantly upwards or downwards.  In the evaporation/condensation scenario, generally, blueshifts are observed in high-temperature lines, while redshifts are a feature of chromospheric and transition-region lines \citep{Milligan2009}. Given that the Lyman lines are predominantly chromospheric \citep{Vernazza1981,Rubiodacosta2009}, we had expected to observe downflows in these lines. However, we observe upflows in three of our flares. 
 
To generate evaporative upflows in low-temperature lines would require either very gentle heating of the mid-chromosphere, where the Lyman lines form already in the quiet Sun, in a very gentle evaporation scenario \citep{Zarro1988, Milligan2006, Battaglia2015}.  This is not likely for any of the strong flares in our sample. An alternative explanation could invoke the presence of a cool, neutral hydrogen layer pushed upward by an underlying, hotter plasma, requiring heating primarily deep in the chromosphere. 

Considering options other than those related to chromospheric evaporation, the three events showing blueshifts have evidence from imaging for flows or ejecta. Our AIA analysis has shown a prominent ejection feature in SOL2011-03-07T19:46 (Figures \ref{Fig9}a \& \ref{Fig9}b), which could explain its observed upflows. A previous analysis found the failed ejection of a small filament in SOL2011-11-03T20:20, though it is surprising that such a small feature would lead to a detectable velocity signature. SOL2012-03-07T00:07 was a major eruptive flare so blueshift in the Lyman lines would be unsurprising. The blueshifted emission could result from emission produced by, or scattering of chromospheric emission from, the erupting material \citep{2012A&A...537A.100L}

In the events showing redshift there may be some evidence for inflowing material or loop contraction in the core of the active region. However, the momentum-balancing chromospheric condensation counterpart to evaporative upflows should also be important.
 Work is currently ongoing to model the Lyman lines using the RADYN code of \citet{Carlsson1997}, with preliminary findings showing the possibility of both upflows and downflows in the Lyman lines in response to the injection of a single electron beam event. A full understanding of the response of the Lyman lines during a flare will require both observations and modelling.

While a physical picture for the flow direction of the Lyman lines remains challenging, we have nonetheless demonstrated that it is possible to measure plasma speeds of a few tens of km s$^{-1}$ with EVE.  On average, the hydrogen Lyman lines tend to display speeds between 20-30 km s$^{-1}$ during flares, while \ion{C}{iii} tends to be emitted by plasma moving a bit faster, averaging 30 km s$^{-1}$. These speeds do not seem implausible for chromospheric lines. \citet{Chaeetal1998} associate chromospheric motions of H$\alpha$ with speeds of 15-30 km s$^{-1}$, while two chromospheric lines (H$\alpha$ and \ion{He}{i}) studied by \citet{Kamio2005}  exhibit speeds of 7 km s$^{-1}$ and 24 km s$^{-1}$. 
 
The high variability in the speeds obtained using the ``flare-excess" data, highlights the need to compare with those obtained from ``Sun-as-a-star" measurements in order to verify the trends we observe. The low irradiance of the ``flare-excess" data for analysing and fitting the Lyman-lines is a key obstacle in this study.  The focus of future work will be on modelling the Lyman lines using radiation hydrodynamic simulations in order to understand what speeds we should expect to observe, and in applying the same methods to future flares we observe.

\section*{Acknowledgements}
The authors thank the anonymous referee for their helpful and insightful comments, which have prompted additions and improvement to the manuscript.  The authors are also grateful to the EVE team for making the data freely avaliable. We acknowledge support from an STFC research studentship, and also STFC grants ST/I001808/1 and ST/L000741/1. The research leading to these results has received funding from the European Community's Seventh Framework Programme (FP7/2007-2013) under grant agreement no. 606862 (F-CHROMA). 

\bibliographystyle{aa}
\bibliography{bibliography}

\end{document}